\documentclass[twocolumn,english]{revtex4}
\usepackage[T1]{fontenc}
\usepackage[latin1]{inputenc}
\usepackage{float}
\usepackage{graphicx}
\makeatletter

\usepackage{amsmath}
\usepackage{amssymb}
\usepackage{babel}
\makeatother

\begin{document}

\title{Universal scaling in transport out of equilibrium through a single quantum dot using the noncrossing approximation}
\author{P.~Roura-Bas}
\affiliation{Centro At\'omico Constituyentes, Comisi\'on Nacional de Energ\'ia At\'omica, Buenos Aires, Argentina}

\email{roura@tandar.cnea.gov.ar}

\begin{abstract}

The universal scaling behavior is studied for nonequilibrium transport through a quantum dot. To describe the dot we use the standard Anderson impurity model and use the non-equilibrium non-crossing approximation in the limit of infinite Coulomb repulsion. 
After solving de Hamiltonian, we calculate the conductance through the system as a function of temperature $T$ and bias voltage $V$ in the Kondo and in the mixed valent regime. 
We obtain a good scaling function in both regimes. In particular, in the mixed valent regime, we find excellent agreement with recent experiments and previous theoretical works.

\end{abstract}

\pacs{73.63.Kv, 72.15.Qm, 75.20.Hr, 73.23.Hk}
\maketitle

\section{Introduction} 
Scaling laws and critical phenomena are intertwined concepts through renormalization-group (RG) arguments as scaling ideas describe how physical quantities change under RG transformations. Scaling of physical quantities is particularly interesting because they reveal \textit{universal} behaviour within certain global constraints. The power of universal scaling functions is that they associate seemingly different physical phenomena which shown similar behaviour when properly scaled with the aid of some parameters. Specifically, zero-bias conductance through a single channel quantum dot (QD) coupled to a reservoir of conduction electrons as a function of temperature can be accurately fitted with an empirical formula extracted from the numerical renormalization-group calculations, $G_E(T)$ \cite{G_E}. This is due to the fact that at low temperatures and in the Kondo regime there is only one relevant energy scale, the Kondo temperature, $T_K$. Experimental realizations of QD devices have reported a successful match between the obtained zero-bias conductance, appropriately scaled by $T_K$, and the empirical formula $G_E(T)$ \cite{gold,yu,roch}. 

When a finite bias voltage $V$ is applied between the leads coupled to the QD,  
the system is no longer in equilibrium and a new energy scale emerges, $eV$ (electron charge times voltage). 
It is, then, important to restore the corresponding universal scaling behavior. From theoretical considerations on the single QD, it is expected that conductance as a function of temperature and bias voltage, $G(T,V)$, may be described by a scaling relation. This scaling function has a quadratic temperature and voltage dependence, \cite{oguri}, in the low range of $T$ and $eV$ compared to $T_K$. 

In a series of experiments in a spin-$1/2$ Kondo dot, GaAs, Grobis \textit{et al.} \cite{GaAs} measured the low temperature and low bias transport properties obtaining 
a quadratic power law in agreement with theoretical predictions. In addition to this, they suggested a universal function to describe the non-equilibrium conductance with the aid of only two scaling parameters even in the range of large temperatures as compared to the Kondo one. The scaling parameters are $T_K$ and the height of the resonant peak in the limit that $T\rightarrow0$, $G_0$.

Recent experiments made by Scott \textit{et al.} \cite{scott} on non-equilibrium transport through single molecule transistor in the Kondo regime have tested the universality of the scaling function of Grobis. The authors also report a quadratic form of the scaling exponents in the low energy range  

On the other hand, by using non-equilibrium renormalized perturbation theory (RPT), Rinc\'on \textit{et al.} \cite{Armando} studied the proposed universal law within the Kondo and the mixed valence regimes. The authors found appropriate scaling relation in both regimes when using the suggested scaling law. In addition, in the case of the mixed valence regime the parameters involved in the scaling formula match with the experimentally obtained for GaAs.

In this work, we study the scaling properties of $G(T,V)$ through a spin-$1/2$ QD by using the non-equilibrium non-crossing approximation. We obtain the conductance  as a function of temperature and bias voltage within the different regimes of the model describing the QD. In what follows we discuss the observed scaling relation for each parameter range. The results for conductance and the extracted scaling parameters are compared with experiments and previous theoretical works \cite{GaAs,scott,Armando}.

\section{Model and method} 

To describe the system in which the QD is coupled to two leads we use the spin $1/2-$ Anderson Hamiltonian:

\begin{eqnarray}\label{anderson}\begin{split}
H=& \sum_{k\nu\sigma}\varepsilon_{\nu\sigma} c^{\dagger}_{k\nu\sigma} 
     c_{k\nu\sigma} + \\
  & \sum_{\sigma}E_{d} n_{d\sigma} + U n_{d\uparrow}n_{d\downarrow}+ \\
  & \sum_{k\nu\sigma}V_{k\nu} d^{\dagger}_{\sigma} c_{k\nu\sigma}+ H.c. .\\
\end{split}\end{eqnarray}

Here $c^{\dagger}_{k\nu\sigma}$ creates a conduction electron with momentum $k$ and spin $\sigma$ in the lead $\nu$ where $\nu=L,R$ labels the left and right leads. $d^{\dagger}_{\sigma}$ creates an electron in the dot and $n_{d\sigma}$ is the number operator for a given spin in the QD. The energy of a single electron in the dot and the Coulomb repulsion are represented by $E_{d}$ and $U$ respectively. The hybridizations between the leads and the QD are given by $V_{k\nu}$ and the lead-dot coupling strengths are

\begin{equation}\label{hybridization}
\Delta_{\nu}(\omega)\equiv\pi\sum_{\kappa\nu} V_{\kappa\nu}^{2}\delta(\omega - \epsilon_{\nu\sigma}),
\end{equation}

which we consider to be step functions  $\Delta_{\nu}(\omega)=\Delta_{\nu} \theta(D-\vert\omega\vert)$ with an energy cut-off $D$ of the order of several times $\Delta_{\nu}$. 
 
The bias voltage $V$ is directly related with the difference between the chemical potentials $\mu_L$ and $\mu_R$ of the leads, $eV=\mu_L-\mu_R$ and we set $\mu_L=eV\Delta_{R}/\Delta$ where $\Delta=\Delta_{L}+\Delta_{R}$ represents the total coupling between the leads and the QD. In this way, in the case of symmetric  couplings, $\Delta_{L}=\Delta_{R}$, the chemical potentials on both sides of the dot are related by $\mu_L=-\mu_R=eV/2$.

In a generic experimental configuration there are no restrictions on the relative couplings to the leads and the current through the dot can be expressed in terms of the lesser and greater Green functions of the QD \cite{meir}. Otherwise, in the case of a proportional coupling, $\Delta_L=\lambda\Delta_R$, the current through the dot may be obtained solely in terms of the QD spectral function \cite{win}

\begin{equation}\label{current}
I(V) =  \frac{e}{\hbar} A \Delta \int d\omega \rho_d(\omega) 
        \left(  f(\omega-\mu_L) - f(\omega-\mu_R)\right) ,
\end{equation}

where $A=4\Delta_L\Delta_R/\Delta^{2}$ represents the asymmetry factor of the couplings between both leads and the dot. The function $f(\omega)=1/(1+e^{\beta\omega})$ is the Fermi distribution and the spectral function $\rho_d(\omega)$ is given by $\rho_d(\omega)=-Im G^{r}_{d\sigma}(\omega)/\pi$, where $G^{r}_{d\sigma}(\omega)$ is the retarded Green function of the dot. 

In this work, we calculate $\rho_d(\omega)$ using the non crossing approximation (NCA) method in its strong coupling (SC) limit $U\rightarrow\infty$. The NCA is a self-consistent conserving scheme in which the local operator $d_\sigma$ is expressed by a combination of auxiliary boson and pseudo-Fermion particles \cite{bickers,bickers_2}. It yields an accurate quantitative description of the equilibrium single channel Anderson model down to low temperatures, although it does not capture the correct Fermi-liquid behavior. As an additional advantage of the method, the NCA can be generalized out of equilibrium in a straightforward way \cite{win,hettler} and it results in a reliable approximation for quantities involving the $\rho_d(\omega)$ such as the conductance as a function of temperature and bias voltage \cite{roura_1,roura_2}. In the numerical procedure to solve the NCA equations, we have used a set of self-adjusting meshes (rather than the fixed one used in Ref. \cite{hettler}) to describe the spectral densities and lesser Green's functions of the auxiliary particles. The procedure guarantees the resolution of the sets of integral equations for the auxiliary particles self-energies to a high degree of accuracy. 

As we mentioned in the previous section, the purpose of this work is to test the experimentally observed scaling relation\cite{GaAs,scott} and to provide a theoretical complement to the recent RPT study \cite{Armando}. 

The proposed universal scaling relation for the conductance as a function of temperature and bias voltage is given by the following expression

\begin{equation}\label{ley_extendida}
\frac{G(T,V)}{G_E(T)} \simeq 1-\frac{\alpha c_T \left( eV/\kappa T_K\right)^{2} }
 {1+(\gamma/\alpha - 1 )c_T\left( T/T_K\right)^{2}},
\end{equation}

where $G_E(T)$ is the empirical formula derived from a fit to numerical renormalization group conductance calculations

\begin{equation}\label{ley_empirica}
G_E(T)=\frac{G_0}{\left[ 1+ \left( 2^{1/s} - 1 \right)
                      \left( T/T_K\right)^{2}   \right] ^{s}},
\end{equation}

and $s=0.21$ for a spin-$\frac{1}{2}$ impurity. The constant $c_T=5.49$ is given by the low temperature expansion of Eqs. (\ref{ley_extendida}) and (\ref{ley_empirica}). The constants $\alpha$ and $\gamma$ represent the curvature at $T\rightarrow0$ and the evolution with increasing temperature of the resonant peak of the $G(T,V)$, respectively. In a more general case when $\gamma=\gamma(T)\sim1/T$, the low temperature limit of the Ansatz given by Eq. (\ref{ley_extendida}) agrees with the scaling Ansatz  discussed by Hettler \textit{et al.} \cite{hettler} ( see Fig. 3b). This means that the proposal Eq. (\ref{ley_extendida}) in which $\gamma$ is not temperature-dependent represents the simplest approximation for $\gamma(T)$.

The empirical law, $G_E(T)$, allows a definition of $T_K$ as the temperature at which the equilibrium conductance is equal to half its zero temperature value, $G(T_K)=G_0/2$. 

The scaling coefficients $\alpha$ and $\gamma$ are obtained and discussed in the next section within different regimes of the Anderson Hamiltonian given in the Eq. (\ref{anderson}).

\section{Numerical Results}

In the strong coupling limit, $U\rightarrow\infty$ forbids double occupancy of the dot, therefore the number of electrons in the QD is restricted to be $\langle n \rangle \leq 1 $. In this limit there are tree different regimes of the QD, the Kondo-Regime (KR), the Mixed-Valence regime (MV) and the Empty-Orbital one (EO). 
In terms of the occupancy of the dot, the KR is characterized by $\langle n_{KR} \rangle \sim 1 $. On the other hand, in the MV some charge fluctuations are allowed and therefore $\langle n_{MV} \rangle < 1 $.  

In the following we present the numerical results for the differential conductance as a function of temperature and bias voltage within the Kondo and mixed valence regimes in the $U\rightarrow\infty$ limit. We focus on the case of symmetric coupling, $A=1$, between the lead and the QD. This assumption is justified by the experimental setup used in the measurement of conductances for GaAS, that keep the two couplings nearly equal, \cite{GaAs}.

From now on, the total coupling $\Delta$ is taken as the unity of energy and we set the bandwidth to be $D=10\Delta$.

\begin{figure}[!ht]
\includegraphics[clip,scale=.45]{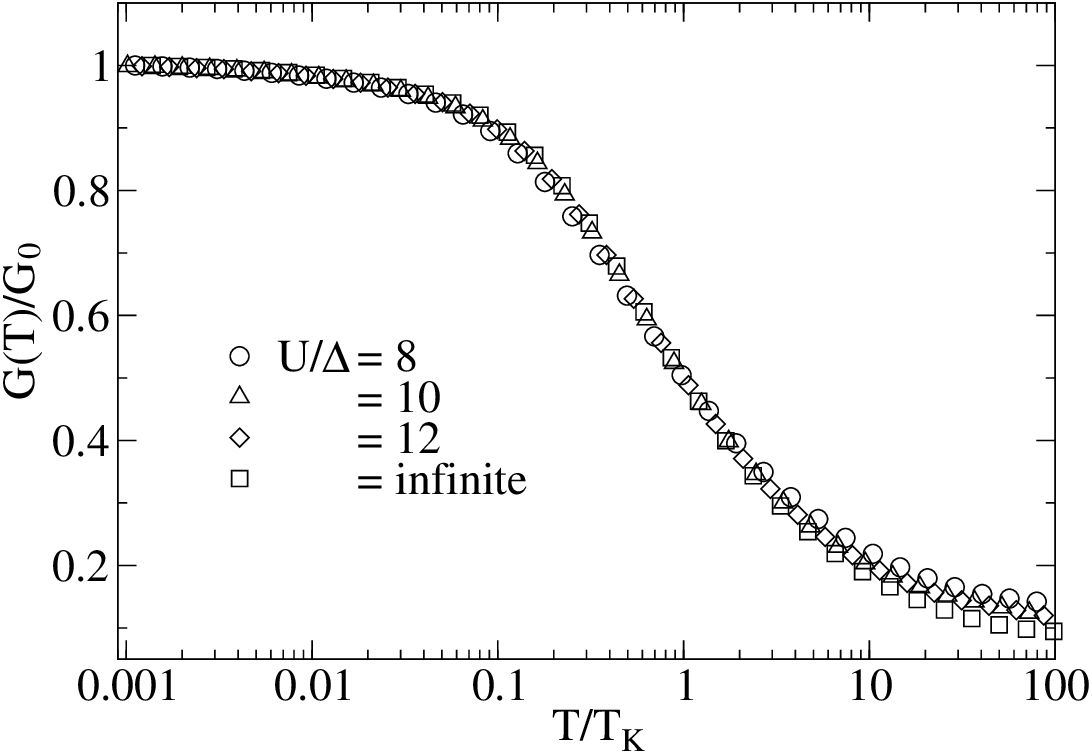}
\caption{Equilibrium conductance versus $T/T_K$ for several values of U. The valence of the QD is set around of 0.95. $T_K$ is defined as the temperature such that $G(T_K)\equiv G_0/2$.}
\label{compara_U}
\end{figure}

The reader might wonder if a finite $U$ expected in realistic systems affects our conclusions. We show that in the Kondo regime, 
to take infinite $U$ is not an essential approximation. Below we will show that in the mixed valence regime, our results coincide 
with those of alternative treatments developed for small $U$. 
In order to clarify the validity of the U infinite approximation concerning the scaling properties we compare in Fig.(\ref{compara_U}) the equilibrium conductance $G(T)/G_0$ as a function of $T/T_K$, keeping constant the valence of the QD, for several values of the Coulomb repulsion U. To make a valid comparison, we use the finite \cite{U_finito} and infinite versions of the NCA well inside the Kondo regime with an occupation of the QD within the range $0.95 < \langle n \rangle < 0.96$.

In the numerical evaluations we normalized the conductance to $G_0\equiv G(T_0,0)$, where $T_0$ is the lowest temperature that can be reached within our NCA calculations ($T_0\sim0.005T_K - 0.01 T_K$).

From Fig.(\ref{compara_U}) it is clear that within the Kondo regime, in which the charge fluctuations are frozen, the only relevant scale is given by the Kondo temperature, $T_K$. In Ref.\cite{gerace}, Gerace \textit{et al.} performs a detailed calculations of scaling behaviour with finite U values within NCA.

As we mentioned in the previous section, in what follows, we obtain the conductance through the QD as a function of temperature and bias voltage within the strong coupling limit $U\rightarrow\infty$ in the Kondo regime and mixed valence one.

Within the NCA these regimes are schematically indicated in the inset of the Fig.\ref{scaling_T}. As it is shown the cases $E_d \leq -3\Delta$ and $-2\Delta \leq E_d \leq 0$ can be considered as describing the KR and MV regimes, respectively.

In order to obtain the differential conductance $G(V) = dI(V)/dV$, we perform a numerical differentiation of the current in Eq. (\ref{current}) for a given $T$. 
Having calculated $T_K$ from the empirical law $G_E(T)$, the value of $\alpha$ in the universal scaling function, Eq. (\ref{ley_extendida}), can be determined from its low temperature expansion  

\begin{equation}\label{ley_alpha}
G(T_0,V)/G_0 \sim  1 - \alpha c_T \left( eV / \kappa T_K\right)^{2} .
\end{equation}

Notice that from RPT calculations in the strong coupling limit $U\rightarrow\infty$, the value of the $\alpha$ parameter does not depend on the asymmetry $A$. This follows from Eq. (\ref{rpt}), in which the strong coupling limit is given by the renormalized parameter $\widetilde{u}=\widetilde{U}/(\pi\widetilde{\Delta})\rightarrow1$ \cite{alpha_T}.   
\begin{eqnarray}
\frac{G}{G_{0}} &\simeq &1-\frac{\pi^{2}(1+2\widetilde{u}^{2})}{3}\left(
\frac{kT}{\widetilde{\Delta}}\right) ^{2} \notag \\
&&-\frac{4-3A+(2+3A)\widetilde{u}^{2}}{4}\left( \frac{eV}{\widetilde{\Delta}}
\right) ^{2}.  \label{rpt}
\end{eqnarray}

Once $\alpha$ is obtained, the $\gamma$ parameter is found by plotting the scaled conductance $(1-G(T,V)/G(T,0))/\alpha_v$ versus $(eV/\kappa T_K)^{2}$ where 

\begin{equation}\label{alpha_v}
\alpha_v = \frac{\alpha c_T}{1+(\gamma/\alpha - 1 )c_T\left( T/T_K\right)^{2}}.
\end{equation}

\subsection{Kondo regime}

\begin{figure}[!ht]
\includegraphics[clip,scale=.45]{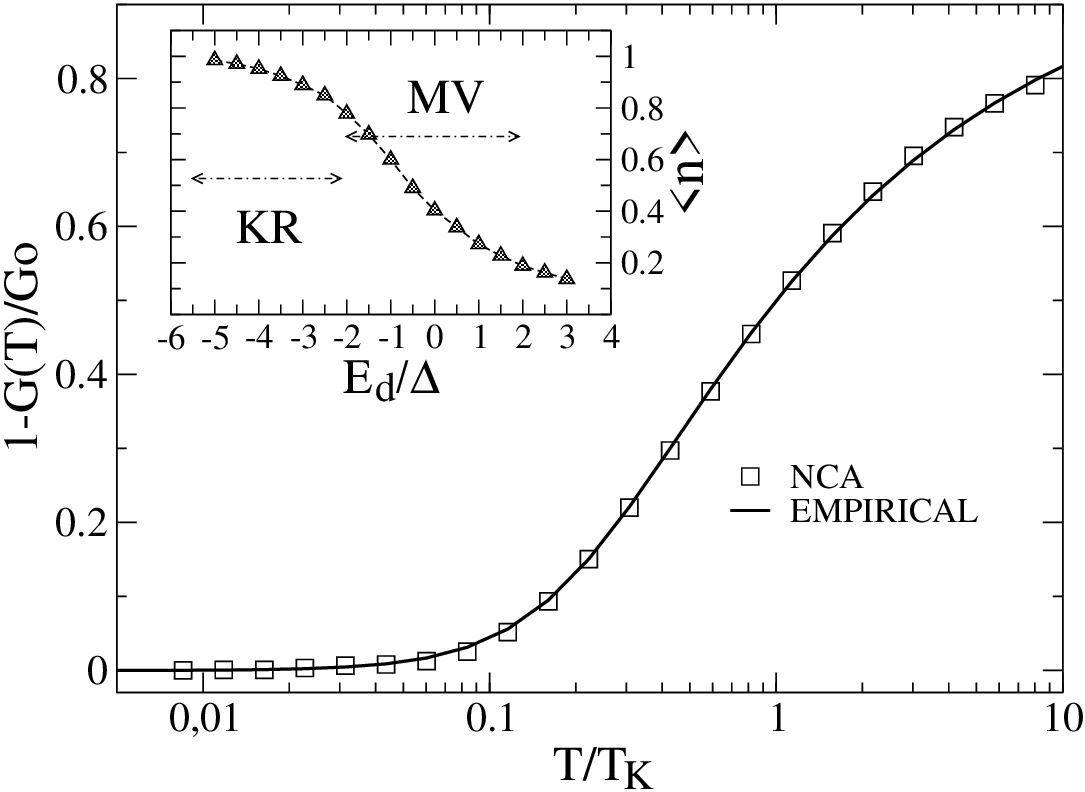}
\caption{Equilibrium conductance versus empirical law $G_E(T)$ as a function of temperature within the KR regime, $E_d=-3\Delta$. The squares represent the NCA conductance while the solid line is for the empirical law defining $T_K=0.038\Delta$. In the inset the different regimes are schematically indicated as a function of $E_d/\Delta$.}
\label{scaling_T}
\end{figure}

We begin analyzing the case in which $E_d=-3\Delta$ and the QD is symmetric connected to the leads, $\Delta_{L}=\Delta_{R}$. We obtain $T_K$ by a fit the NCA linear response conductance ($V=0$) with the empirical law $G_E(T)$. In Fig.\ref{scaling_T}, we compare the empirical law with the NCA result. The obtained conductance follows the empirical law very well in the whole range of temperatures and fixes the value of the Kondo temperature as $T_K=0.038\Delta$.

\begin{figure}[!ht]
\includegraphics[clip,scale=.45]{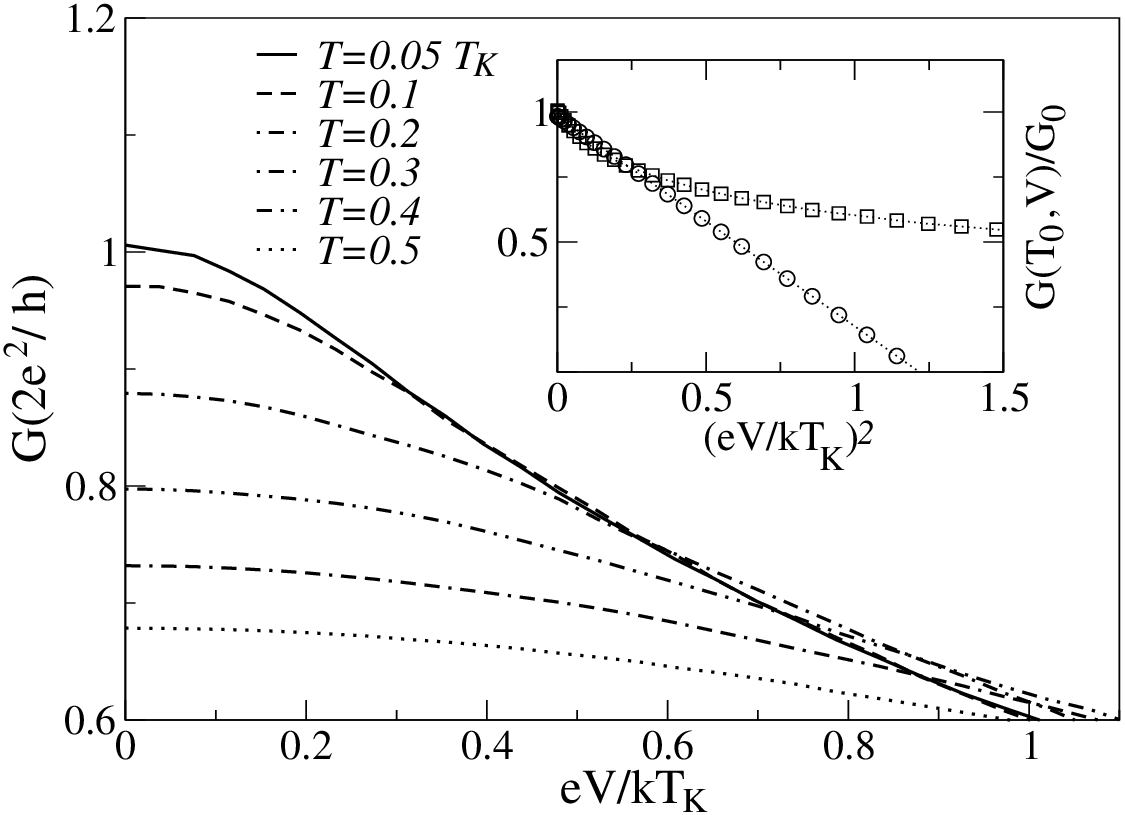}
\caption{ Differential conductance as a function of $eV/\kappa T_K$ for several temperatures below $T_K$ and for $Ed=-3\Delta$. The inset shows the linear regression curve that fitted $G(T,V)$ very well up to $(eV/\kappa T_K)^{2}\sim0.2$ at $T_0=0.05T_K$. The  obtained value for  $\alpha$ is $\alpha=0.16 \pm 0.01$}
\label{alpha_Ed_3}
\end{figure}

When a finite bias $V$ is applied the differential conductance as a function of the bias voltage exhibit the usual Kondo peak centered at zero bias and it broadens fast as the temperature increases. The obtained $G(T,V)$ are shown in Fig.\ref{alpha_Ed_3} as a function of $eV/\kappa T_K$ for several temperatures below the Kondo one. It is noticed that in the limit of $T\rightarrow0$ the conductance saturates at the usual value of $2e^{2}/h$ for the Kondo regime. 

From Eq. (\ref{ley_alpha}) we extracted the value of $\alpha$ at very low temperatures as compared to $T_K$ ( $0.01T_K<T<0.05T_K$ ) and over the range $eV/\kappa T_K<0.4$ as in the GaAs experiment. We obtain an average value $\alpha=0.16\pm0.01$. This value of $\alpha$ is consistent with the $\alpha^{SAM}_{SC}=3/(2\pi^{2})=0.152$ predicted for the Anderson model in its strong $U\rightarrow\infty$ limit \cite{Armando}. The inset of Fig.\ref{alpha_Ed_3} shown the fitted conductance for  $T=0.05T_K$.

The values of $\gamma$ for each temperature are obtained from Eq.\ref{alpha_v} and are listed in Table \ref{Tabla1}. From the temperature dispersion, Table \ref{Tabla1}, the average value of $\gamma$ is $\gamma\sim0.9\pm0.2$. The obtained values for $\alpha$ and $\gamma$ are major than the corresponding ones to the experiment.

\begin{table}[ht]
\centering
\caption{Shown are the calculated values for $\alpha_v$ and $\gamma$ as a function of temperature for $E_d=-3\Delta$. Temperature in units of $T_K$.}
\begin{tabular}{c c c}
\hline
\hline
Temperature &   $\alpha_v$   &      $\gamma$  \\ \hline

0.05        &   0.743        &   1.33         \\
0.10        &   0.643        &   0.93         \\
0.20        &   0.402        &   0.87         \\
0.30        &   0.270        &   0.77         \\
0.40        &   0.178        &   0.77         \\
0.50        &   0.132        &   0.73         \\

\hline
\hline
\end{tabular}
\label{Tabla1}
\end{table}

\begin{figure}[!ht]
\includegraphics[clip,scale=.45]{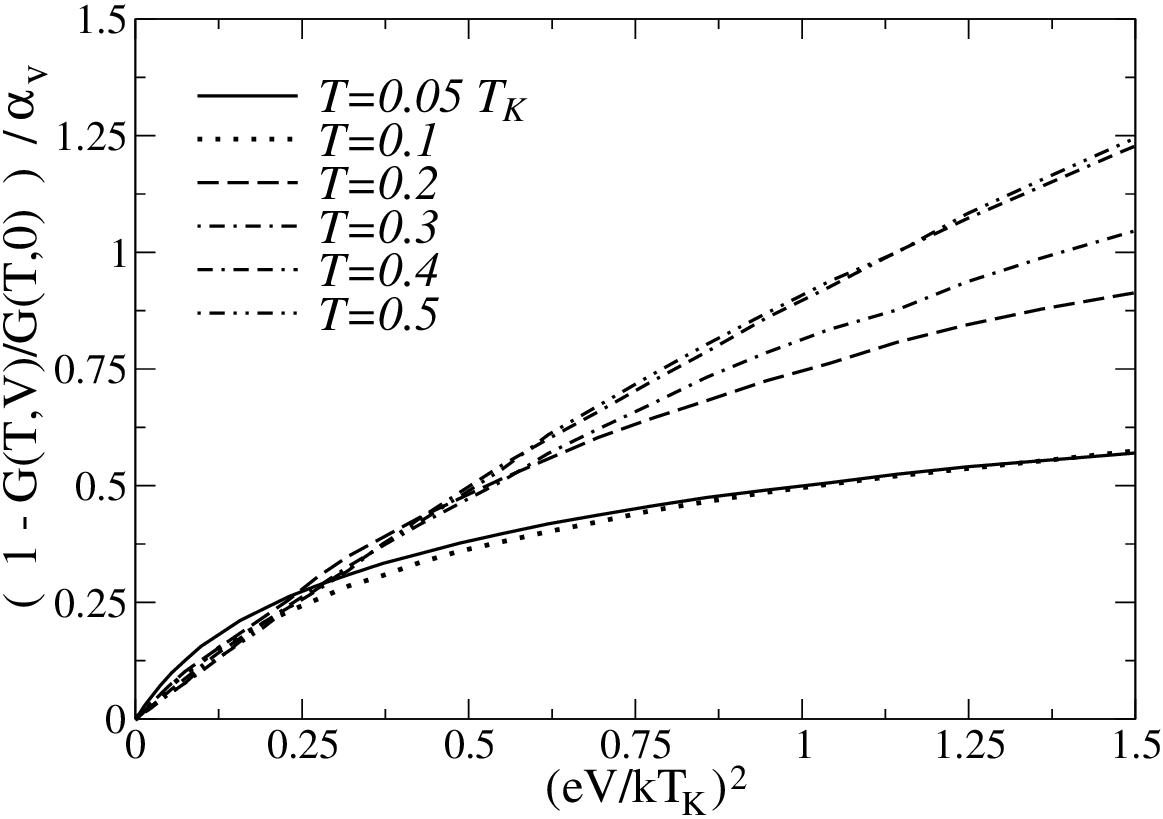}
\caption{ Scaled conductance $(1-G(T,V)/G(T,0))/\alpha_v$ as a function of $(eV/\kappa T_K)^{2}$ for several temperatures in the case of $E_d=-3\Delta$. The temperatures are given in units of $T_K$. A collapse of all curves into a linear $y=x$ is obtained.}
\label{scaling_VT_Ed_3}
\end{figure}

In Fig.\ref{scaling_VT_Ed_3}, we plot the scaled conductance $(1-G(T,V)/G(T,0))/\alpha_v$ as a function of $(eV/\kappa T_K)^{2}$ for several temperatures using the average calculated values of $\alpha$ and $\gamma$. The collapse of all curves in a straight line with slope 1 at low voltages might suggest that the Ansatz scaling is appropriate. However, an important point to note is that this collapse require different values of $\gamma$ for different temperatures. The dependence of $\gamma$ with the temperature could indicates the breakdown of the scaling Ansatz  given by Eq. (\ref{ley_extendida}) in the Kondo regime. The decrease of $\gamma$ with increasing temperature, as was mentioned in the previous section, agrees with a more general relation in which $\gamma$ is proportional to $1/T$ \cite{hettler}. 
This deviation of the Ansatz scaling due to the temperature dispersion  of $\gamma$ was also found by using RPT (Ref. \cite{Armando}). Besides, our results agree with the obtained by using RPT, $\alpha=0.151$ and $\gamma=0.8$, given in Ref. \cite{Armando}.

\subsection{Mixed valence regime}

\begin{figure}[!ht]
\includegraphics[clip,scale=.3]{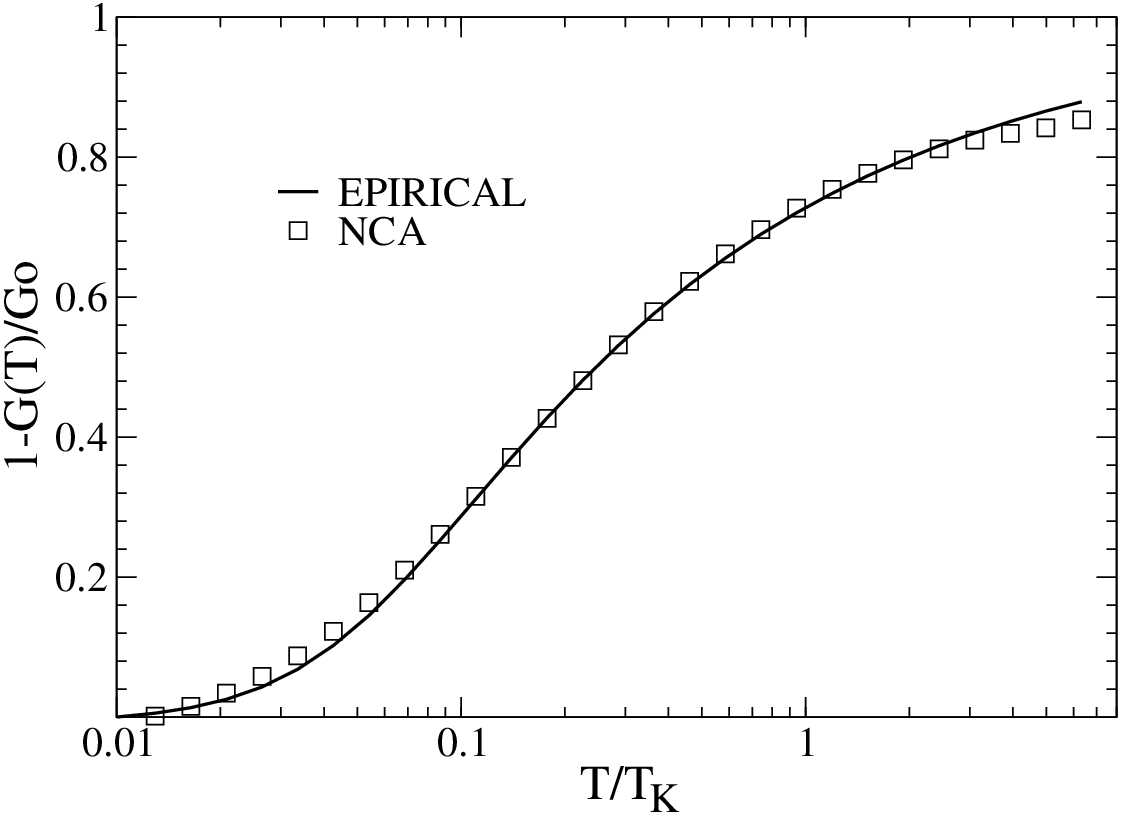}
\caption{Equilibrium conductance versus empirical law $G_E(T)$ as a function of temperature within the MV regime, $E_d=-2\Delta$. The squares represent the NCA conductance while the solid line is for the empirical law defining $T_K=0.158\Delta$.}
\label{scaling_2_T}
\end{figure}

We now focus on the nonequilibrium transport through the dot for $E_d=-2\Delta$. In this regime of the parameters the occupancy in the QD is $\langle n \rangle \sim 0.8 $ (inset of Fig.\ref{scaling_T}) so some charge fluctuations are allowed.
From the conductance at zero bias and its fitting with the $G_E(T)$, the corresponding Kondo temperature is found to be $T_K=0.158\Delta$. As it is shown in Fig.\ref{scaling_2_T} the fit of the NCA conductance with the empirical law is very good in the whole range of temperature. 

\begin{figure}[!ht]
\includegraphics[clip,scale=.45]{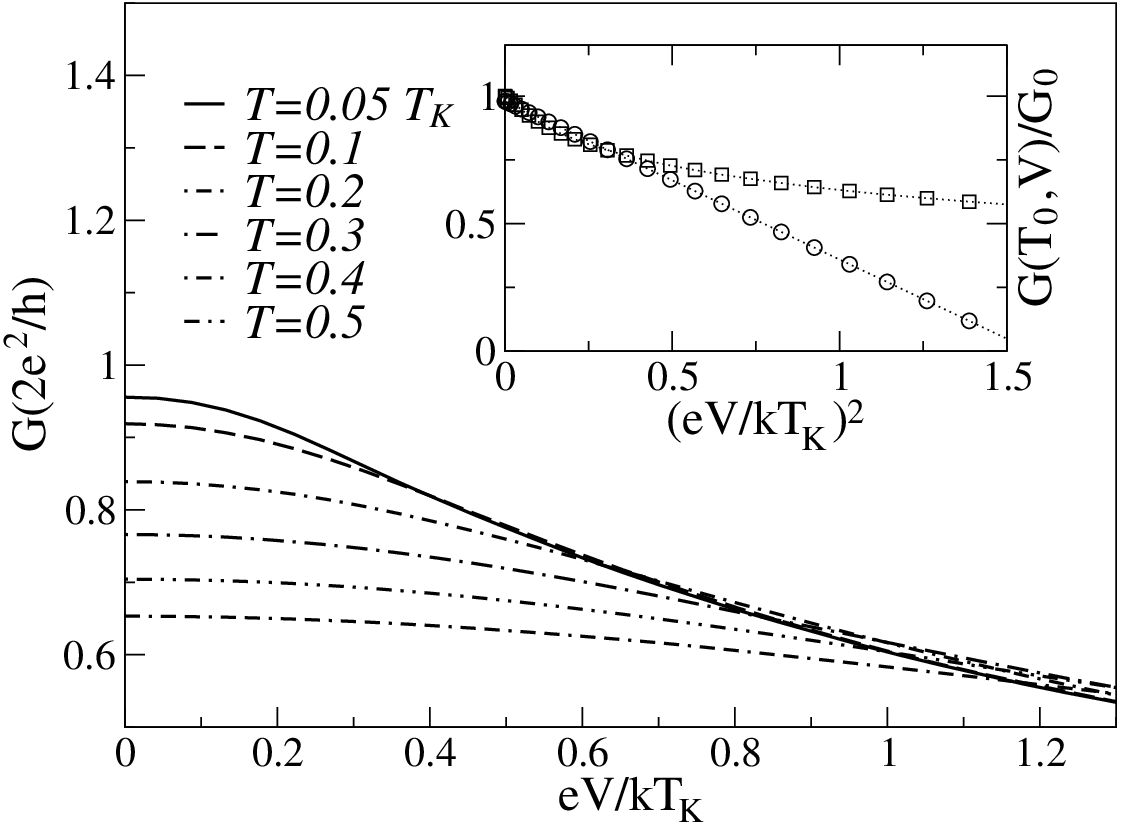}
\caption{ Differential conductance as a function of $eV/\kappa T_K$ for several temperatures below $T_K$ for $E_d=-2\Delta$. The inset shows the linear regression curve that fitted $G(T,V)$ very well up to $(eV/\kappa T_K)^{2}\sim0.3$ at $T=0.05T_K$. The  obtained value for  $\alpha$ is $\alpha=0.11 \pm 0.01$.}
\label{alpha_Ed_2}
\end{figure}

Figure \ref{alpha_Ed_2} shows the temperature evolution of the conductance as a function of the  applied bias. As it can be seen from a comparison between Figs.\ref{alpha_Ed_3} and \ref{alpha_Ed_2}, the curvature of the conductance at $T\rightarrow0$ is smoother in the present case than for $E_d=-3\Delta$. It is noticed that the conductance saturates at a value less than the corresponding to the Kondo regime ($2e^{2}/h$) due to the allowed charge fluctuations. 

Following the procedure already described in the previous section, from the $G(T,V)$ at $T=T_0$ we obtain $\alpha=0.11\pm0.01$. This value of $\alpha$ is in a surprising agreement with the one obtained by Grobis \textit{et al.} in the GaAs experiment \cite{GaAs}. Moreover, this is qualitatively the same feature that was found within the RPT \cite{Armando}.

The $\gamma$ values within this regime, and more important their dispersion are, also reduced in comparison with the respectives obtained for the KR. Table \ref{Table2} shows the values of $\alpha_v$ and $\gamma$ from the fit of conductance for each temperature. The average value of $\gamma$ from the temperature dispersion is $\gamma\sim0.48\pm0.03$, in excellent agreement with the experimental value $\gamma\sim0.5\pm0.1$, \cite{GaAs}. This value agree with that found by Rinc\'on \textit{et al.} in this regime although they used a symmetric Anderson model, $E_d=-U/2$, and finite $U$ in contrast to our case.

\begin{table}[ht]
\centering
\caption{Shown are the calculated values for $\alpha_v$ and $\gamma$ as a function of temperature for $E_d=-2\Delta$. Temperature in units of $T_K$.}
\begin{tabular}{c c c}
\hline
\hline
Temperature &   $\alpha_v$   &      $\gamma$  \\ \hline

0.05        &   0.575        &   0.514        \\
0.10        &   0.505        &   0.502        \\
0.20        &   0.340        &   0.498        \\
0.30        &   0.225        &   0.485        \\
0.40        &   0.158        &   0.462        \\
0.50        &   0.115        &   0.452        \\

\hline
\hline
\end{tabular}
\label{Table2}
\end{table}

\begin{figure}[!ht]
\includegraphics[clip,scale=0.45]{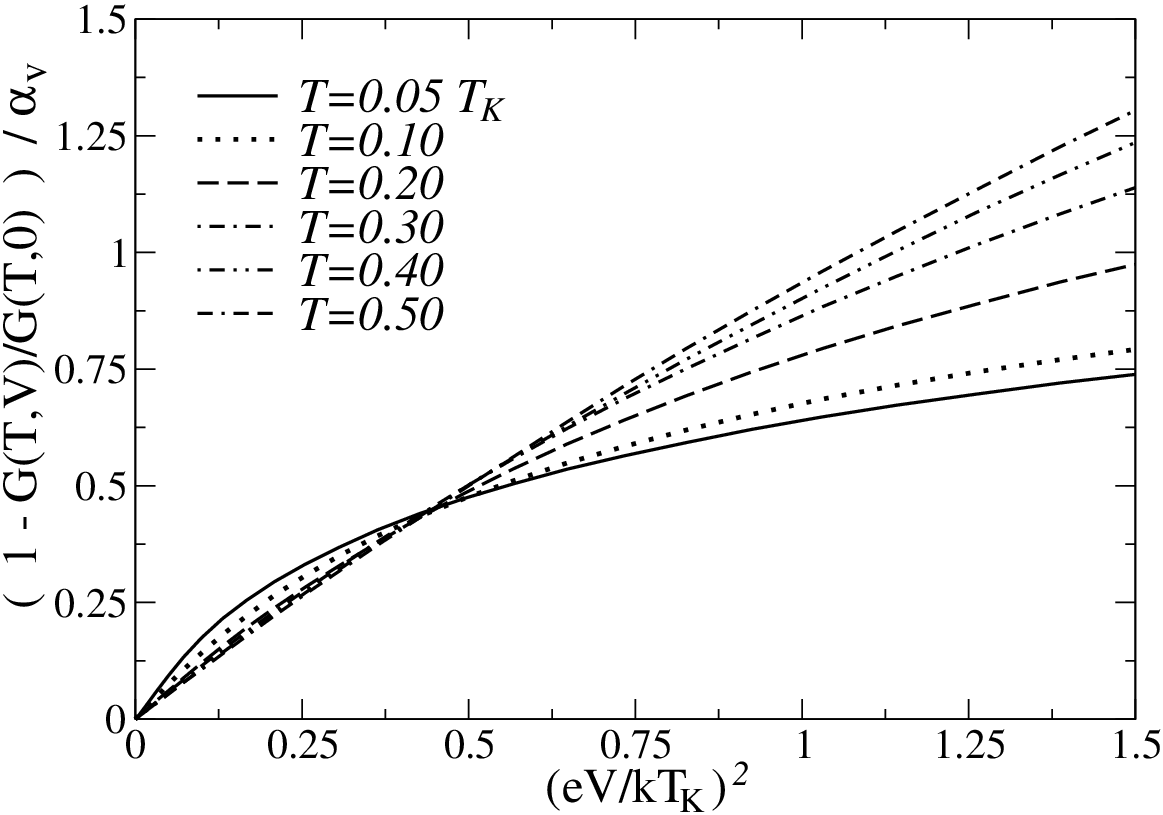}
\caption{Scaled conductance $(1-G(T,V)/G(T,0))/\alpha_v$ as a function of $(eV/\kappa T_K)^{2}$ for several temperatures in the case of $E_d=-2\Delta$. The temperatures are given in units of $T_K$. A collapse of all curves into a linear $y=x$ is obtained.}
\label{scaling_VT_Ed_2}
\end{figure}

Fig.\ref{scaling_VT_Ed_2} shows the scaled conductance $(1-G(T,V)/G(T,0))/\alpha_v$ versus $(eV/\kappa T_K)^{2}$ for several temperatures using the average calculated values of $\alpha$ and $\gamma$. We notice that the Ansatz  for the universal scaling relation works very well up to at least $(eV/\kappa T_K)^{2}\lesssim0.5$ and for $0.05<T/T_K\lesssim0.5$. As in the case of KR it can be seen that all curves collapse into the linear $y=x$. 

\section{Conclusions}

In the previous sections we analyzed the Ansatz  of Grobis \textit{et al.} for the universal scaling of the conductance out of equilibrium through a quantum dot described for a spin $1/2-$ Anderson model. We performed a non-equilibrium non-crossing approximation study of the conductance within the Kondo and Mixed Valence regimes. In both regimes we obtain an appropriate scaling relations and our results agree with the previous ones by using the non-equilibrium renormalized perturbation theory. However, in the Kondo regime, different values of $\gamma$ for different temperatures $T$ should be used. This suggests that the fitting formula is inappropriate, at least in the Kondo regime, and the dependence of $\gamma$ with temperature should be considered.

This is the same observation obtained by Rinc\'on \textit{et al.} \cite{Armando} even the method and the model are quite different that the ones used in this work.
Both techniques, non-crossing approximation and renormalized perturbation theory, are suitable methods for study the transport properties out of equilibrium for the Anderson model. The agreement represents a good benchmark between both of them. 

The quasi-constant value of $\gamma$ in the mixed valence regime indicates that the Ansatz  given by Eq. (\ref{ley_extendida}) is an appropriate scaling relation in this regime. 
The scaling parameters depend on the regime of the dot. In particular, if valence fluctuations are allowed, the parameters are in excellent match with the corresponding ones obtained in experiments on GaAs QD's.  
Furthermore, this agreement with the experimental values of the $\alpha$ and $\gamma$ parameters suggests that the system in the experiments of Grobis \textit{et al.} presents some valence fluctuations characteristic of the mixed valence regime. 

\section{ACKNOWLEDGMENTS}

I thank A. A. Aligia for helpful conversations. I also thank Ana M. Llois and A. A. Aligia for a critical reading of the manuscript. 
This work was done in the framework of projects UBACyT X115, PICT-33304, PME117 and PIP-CONICET 6016.


\begin{thebibliography}{99}

\bibitem{G_E} T. A. Costi, Phys. Rev. Lett. \textbf{85}, 1504 (2000).

\bibitem{gold} D. Goldhaber-Gordon, H. Shtrikman, D. Mahalu, D. Abusch-Magder, U. Meirav and M. A. Kastner, Nature \textbf{391}, 156 (1998).

\bibitem{yu} L. H. Yu, Z. K. Keane, J. W. Ciszek, L. Cheng, J. M. Tour, T. Baruah, M. R. Pederson and D. Natelson, Phys. Rev. Lett. \textbf{95}, 256803 (2005).

\bibitem{roch} N. Roch, S. Florens, V. Bouchiat, W. Wernsdorfer and F. Balestro, Nature \textbf{453}, 633 (2008).

\bibitem{oguri} A. Oguri, J. Phys. Soc. Jpn. \textbf{74}, 110 (2005).

\bibitem{GaAs} M. Grobis, I. G. Rau, R. M. Potok, H. Shtrikman and D. Goldhaber-Gordon, Phys. Rev. Lett. \textbf{100}, 246601 (2008).

\bibitem{scott} G. D. Scott, Z. K. Keane, J. W. Ciszek, J. M. Tour and D. Natelson,  Phys. Rev. B \textbf{79} 165413 (2009).

\bibitem{Armando} Juli\'{a}n Rinc\'{o}n, A. A. Aligia, and K. Hallberg, Phys. Rev. B \textbf{79}, 121301(R) (2009); Phys. Rev. B
\textbf{80}, 079902(E) (2009); Phys. Rev. B \textbf{81}, 039901(E) (2010).

\bibitem{meir} Y. Meir and N. S. Wingreen, Phys. Rev. Lett. \textbf{68},
2512 (1992).

\bibitem{win} N. S. Wingreen and Y. Meir, Phys. Rev. B \textbf{49}, 11040
(1994).

\bibitem{bickers} N. E. Bickers, D. L. Cox and J. W. Wilkins, Phys. Rev. B \textbf{36}, 2036 (1987).

\bibitem{bickers_2} N. E. Bickers, Rev. Mod. Phys. \textbf{59}, 845 (1987).

\bibitem{hettler} M. Hettler, J. Kroha and S. Hershfield, Phys. Rev. B \textbf{58}, 5649 (1998).

\bibitem{roura_1} P. Roura-Bas and A. A. Aligia, Phys. Rev. B \textbf{80}, 035308 (2009).

\bibitem{roura_2} P. Roura-Bas and A. A. Aligia, J. Phys.: Condens. Matter \textbf{22}, 025602 (2010).

\bibitem{U_finito} \textit{It is noticed that the NCA in its finite U version omits vertex corrections and therefore the Kondo temperature is severely underestimated. However, after scaled by $T_K$, the conductance as a function of temperature follows the universal dependence.}

\bibitem{gerace} D. Gerace, E. Pavarini and L. C. Andreani, Phys. Rev. B \textbf{65}, 155331
(2002).

\bibitem{alpha_T} A. A. Aligia, private communication.



























 
\end{thebibliography}
\end{document}